\documentclass{interact}

\usepackage[doublespacing]{setspace}
\usepackage{xurl}
\usepackage{moreverb,url}
\usepackage[colorlinks,bookmarksopen,bookmarksnumbered,citecolor=red,urlcolor=red]{hyperref}
\usepackage[utf8]{inputenc}   
\usepackage[T1]{fontenc}      
\usepackage{multirow}
\usepackage{placeins}
\usepackage{array}        
\usepackage{longtable}
\usepackage{booktabs}

\usepackage{biblatex}

\begin{document}

\articletype{article}


\title{Open Source Software and Data for Human Service Development: A Case Study on Predicting Housing Instability}

\author{
\name{M.Y. Rodriguez\textsuperscript{a}\thanks{CONTACT M.Y. Rodriguez Email: myr2@buffalo.edu}, E. Dohler\textsuperscript{b}, J. Phillips {c}, M. Villodas {d}; V. Vegara {e}, K. Joseph {a}; A. Wilson {b}}
\affil{\textsuperscript{a}Department of AI and Society, University at Buffalo (SUNY); \textsuperscript{b}School of Social Work, University of North Carolina (Chapel Hill); \textsuperscript{c}School of Social Work, University of Minnesota (Duluth); \textsuperscript{d}School of Social Work, George Mason University}; \textsuperscript{e}Department of Computer Science and Engineering, University at Buffalo (SUNY)}

\maketitle

\begin{abstract}
Open-source data and tools are lauded as essential for replicable and usable social science, though little is known about their use in resource constrained human service provision. This paper examines the challenges and opportunities of open-source tools and data in human service development by using both to forecast “failure to pay” eviction filings in Bronx County, NY. We use zip code level data from the Housing Data Coalition, the American Community Survey 5-year estimates, and DeepMaps Model of the Labor Force to forecast rates through July 2021. We employ multilevel (MLM) and exponential smoothing (ETS) models using the R project for Statistical Computing, an oft used open-source statistical software. We compare our results to what happened during the same period, to illustrate the efficacy of the open-source tools and techniques employed. We argue open-source data and software may facilitate rapid analysis of public data - a much-needed ability in human service intervention development under increasingly constrained resources - but find public data are limited by the information they reliably capture, limiting their utility by a non-trivial margin of error.  The manuscript concludes by considering lessons for human service organizations with limited analytical resources and a vested interest in low-resourced communities. 
\end{abstract}

\begin{keywords}
Open Source, Open Data, Statistical Software
\end{keywords}

\newpage

\section{Introduction}

As technology moves relentlessly forward, it offers an opportunity for relationship oriented fields like human services to reflect on the ways in which technology can help bolster its mission. While contemporary human service scholarship related to technology is focused on AI \parencite{rodriguez_introducing_2024, victor_recommendations_2023,tambe_artificial_2018, patton_chatgpt_2023}, one overlooked area is analysis of the open source data and statistical software (henceforth referred to as 'open source tools') upon which AI and other technologies within human services are built . The term \emph{open source} refers to software and/or data that is free to use, free to modify, and free to distribute \parencite{chopra_decoding_2008, lakhani_how_2003, shepherd_opting_2024, niu_survey_2021}.  While there has been some application of open source software in human service research \parencite{rodriguez_computational_2020, chung_text-mining_2022, auerbach_ssd_2013}, whether and how open source data and software serve the applied research process of human service scholarship, particularly in light of its unique relationship to research \parencite{thyer_what_2001, fraser_intervention_2004, fraser_social_1991, brekke_shaping_2012, trinder_social_1996, sheppard_social_1995}, has not been deeply explored in the literature.

Human service research would be joining a majority of social science fields (applied or academic) by incorporating open source tools into its educational and research repertoire. For example, among the most widely used statistical software in the social sciences are R  \parencite{R-Core}, Python \parencite{van1995python}, and MySQL \parencite{iso_sql} - all of which are open source.  While there are no specific reviews documenting how much social science research uses R, Python or MySQL, a report by GitHub notes Python to be one of the fastest growing languages "driven, in part, by its utility in data science and machine learning". Some scholars have compared across these and other open source software and found them to be complementary rather than mutually exclusive \parencite{hug_statistical_1990, munafo_manifesto_2017, shepherd_opting_2024}. Human service research, by contrast, has not adopted these software in research nor in education across degree programs. The reasons for this lack of adoption appear to be multifaceted \parencite{flaherty_using_2025}.

Beginning with the introduction of statistical software into human service research and practice \parencite{hug_statistical_1990, matheson_innovative_1993, mutschler_factors_1990}, conversations concerning advanced computational analysis in human services research have arguably circumvented the issue of open source tools and methods, excluding related concepts like open access \parencite{bowen_open_2013}, and reproducibility \parencite{yaffe_editorwe_2019}. The current paper engages with this gap formally, examining the opportunities and challenges of using open source tools for humans service application. The paper offers a case study born from opportunities offered by the Covid-19 pandemic.  Specifically, like others,  our team was particularly concerned with the end of federal eviction moratorium (August 26th, 2021) and the impact its end would have on the homeless population. Recognizing that most granular housing data is behind a paywall and that the housing interventions used during the pandemic would be time limited \parencite{benfer2021eviction}, we wondered about the kind of housing analysis we could conduct using publicly available data and open source statistical software only. Our aim was to replicate the conditions human service organizations generally operate under by forecasting the number of "failure to pay" evictions in Bronx County NY using only open source tools and data. We undertook this endeavor for two reasons:

\begin{enumerate}
    \item The Covid-19 pandemic demonstrated the need for rapid, scalable intervention development in response to real-time societal shocks; and, 
    \item A process whereby human service organizations can reliably and replicably undertake forecasting may help triage services more efficiently during future emergency events.
\end{enumerate}

The background section offers an overview of the literature on open source software and tools in human service research and teaching. The method sections offers the context and process by which we undertook the forecasting of failure to pay evictions, as well as detailing the methodological assumptions the open source methods employed herein require researchers to make. We offer the results of our forecasts and compare them to what actually happened during the same time frame. These comparisons offer the basis of our discussion, in which we discuss the opportunities and challenges of using open source tools for human service intervention development. 

\section{Background}

\subsection{What are Open Source tools?}
\parencite{niu_survey_2021} notes that the full definition of open source software comes to us from the Open Source Initiative \parencite{opensource_initiative}, which derived its definition from the Debian Social Contract \parencite{debian}. The open source software definition noted above requires ten conditions to be met before calling something open source. Briefly, these are (quoted from pg.2 of \parencite{niu_survey_2021}): 

 \begin{enumerate}
    \item \emph{Free Redistribution}: The software can be freely given away or sold.
    \item \emph{Source Code}: The source code must either be included or freely obtainable.
    \item \emph{Derived Works}: Redistribution of modifications must be allowed.
    \item \emph{Integrity of The Author’s Source Code}: Licenses may require that modifications are redistributed only as patches.
    \item \emph{No Discrimination against Persons or Groups}: no one can be locked out.
    \item \emph{No Discrimination against Fields of Endeavor}: commercial users cannot be excluded.
    \item \emph{Distribution of License}: The rights attached to the program must apply to all to whom the program is redistributed without the need for execution of an additional license by those parties.
    \item \emph{License Must Not Be Specific to a Product}: the program cannot be licensed only as part of a larger distribution.
    \item \emph{License Must Not Restrict Other Software}: the license cannot insist that any other software it is distributed with must also be open source.
    \item \emph{License Must Be Technology}: Neutral: no click-wrap licenses or other medium-specific ways of accepting the license must be required.
\end{enumerate}

 \parencite{bonaccorsi_why_2003} offers a brief history of open source software in their argument for its sustainability. They note one reason open source software proliferate so easily is the community spirit that births and sustains its continued development. For instance, many open source tools are distributed under the GNU GPL (general public license) copyleft license - meaning that if someone makes a program using code licensed under the copyleft license, they must also use the GNU GPL copyleft license for their project - ensuring the derivative code is always open source. In this way, open source software ensures it reproduces itself freely. The authors note that object oriented programming has also facilitated open source software (indeed, the languages  mentioned earlier are object oriented, such a python and R): 
\begin{quote}
    In the object-oriented framework, a program is no longer a monolithic entity, but is organised into several modules...Inside each program, the combination of modules takes place according to a structured hierarchy of dependence relations, but modules entering at the same level of the system can be developed independently from each other (pg. 1247)
\end{quote}

\subsection{How Open Source Tools Work in Practice}

Since their inception, scholars have investigated one of the principal economic paradoxes of open source tools: they require experts to engage in a great deal of technical work for little to no financial compensation.  Why would experts engage in free labor for products that proliferate so widely?\parencite{baytiyeh_open_2010} conducted a survey of contributors to free/open source software (FOSS) (n=110, 38 paid and 72 volunteers) to learn about the motivating factors for their contributions. Respondents reported three key factors: altruism, the ability to create, and the opportunity to learn. Importantly, receiving formal payment (i.e. a salary) was not an important factor for contributors who reported receiving a salary to work on their FOSS. \parencite{lakhani_how_2003} investigates another aspect of contributions to open source tools: technical support. Examining the open source Apache Web Server software (an open source software used by approximately 25\% of all websites globally), the authors find contributors to open source projects do so primarily because it solves their own problems, such as the need for free software that works well in whatever use case, finding solutions to some software bug, etc. Further, users who predominantly take the role of providing answers to such queries (as in the case of Apache web server tech support) do so because there is relatively little cost to them: 
\begin{quote}
    they [tech support respondents] only posted information they already knew 'off the shelf' - they seldom did new problem-solving or searching in order to provide additional information to a help-seeker (pg 939).
\end{quote}
        
\parencite{dembe_statistical_2011} reviewed 535 articles in order to examine which statistical software was used most often in  humans services research. The authors focused on 3 journals they argue were representative of human services research: Health Services Research, Medical Care, and Medical Care Research and Review (pg. 2). Articles included in the final sample were published between 2007 and 2009 and excluded anything but original research articles. 88\% of the studies within the sample  used either STATA or SAS. Importantly, open source statistical software was subsumed in the other category, reflecting 18.5\% of reported software usage. 
        
   
\subsection{Open Source Data in Human Service Research}

Neither open source data nor software is completely absent from human service scholarship. First, we note the slight distinction between open data and public data. Public data are accessible by researchers who are affiliated with academic institutions through filing an IRB and offering evidence of approval to the data repository. Open data are accessible by anyone with access to where the data are held in repository. Public data may be held by academic institutions, whereas open data are held by organizations that may not be academic in nature, such as governmental bodies. With regard to public data, examples of its use on human service scholarship abound. For example, the Fragile Families and Child Well Being Study (now called the Future of Families and Child Well-being Study \parencite{fragile_families}) is an oft used data source for human service scholarship, particularly dissertations. Several other large scale public datasets, such as the National Child Abuse and Neglect Data Set (NCANDS) \parencite{ncands}, the Panel Study on Income Dynamics (PSID) \parencite{psid} and the American Community Survey (ACS) \parencite{american_community_survey} are also prominent in human service scholarship. Open data  - though more proliferate - are arguably  less often used in a research context: examples of these include municipal sites like NYC open data \parencite{nyc_opendata}, Detroit open data \parencite{detroit_opendata}, and Seattle open data \parencite{seattle_opendata}. Some public projects have used these open data in meaningful ways \parencite{subway_map, justice_map, violence_mapping}, and several of these have resulted in peer reviewed publications  \parencite{deangelis_systemic_2024}. We argue \emph{open data may be well positioned to facilitate human service intervention development due to its free (as in cost) availability.} We investigate this argument in our study below, but first turn to the literature on open source tools in human service practice to understand prior attempts at implementing open source technologies in the intervention development and implementation process. 

\subsection{Open Source Software in Human Service Practice}

Access and education have been noted in prior studies as among the most important factors in how human service professional use technology, broadly construed \parencite{grasso_agency-based_1989}. \parencite{flaherty_using_2025} notes that the usage of analytic software in the field begins with the education received by the professionals conducting the analyses. As such, they conduct a survey of n = 117 faculty teaching research methods in schools of social work to investigate perceived barriers to using the open source software R in these courses. The authors note 'unfamiliarity' and 'complexity of use' as the largest barriers to the implementation of R in these research courses, though it is perhaps ironic they conducted their data analysis using Stata and AtlasTi. 

\parencite{mutschler_factors_1990} conducted a survey of participants of 3 workshops at the University of Michigan's School of Social work concerning technology and human services administration. The workshops were divided as follows: a computer workshop for clinicians, a computer workshop for administrators, and a general (non-technical) workshop for managers.  The respondents are not a random sample, though the exploratory results are informative. Specifically, the authors find workshop participants used computers for administrative, structured tasks more than anything else. For example, 22\% reported using computers for client intake, versus  only 7\%  for decision making. Importantly, 57\%  of all respondents reported viewing the use of computers in human services as negative: with clinicians (65\%) and managers (68\%) in the lead. Though these results are from the inaugural years of computer usage in human services, the following quote illustrates a central obstacle to open source tool proliferation in human services:

\begin{quote}
    "it is frequently reported in the literature that human service professionals lag behind in using the available potential information technology... (Greist et al. 1984; Schoech et al, 1985)" pg. 98
\end{quote}

They note that human service organizations need to invest "a considerable amount of time and manpower" (pg. 99) in order to identify decision processes, rules and requisite information in order to apply new technologies broadly, but decision making tools specifically to activities like program planning, treatment decisions, and outcome evaluation. More work is needed to find the most effective methods to teach human service professionals how to use open source analytical tools.  However, \parencite{mutschler_factors_1990} note user training to be more important than attitudes when seeking to promote technology use (pg. 96). Importantly, human service professionals and scholars are generally trained in social work or a related human service field \parencite{bls_social_human_service_assistants}. Anecdotally, and within the United States, such programs generally use SAS or STATA when training MSW and/or Phd students on statistical analysis. The use of these software seem to be related to the prior training of PhD advisers, prior university licensing agreements, and/or perceptions concerning the initial investment required to make the switch to other software \parencite{flaherty_using_2025}. Recent years have seen some forays into open source tools for teaching research in human services, such as \parencite{flaherty_using_2025, perron_teaching_2022, auerbach_ssd_2013}.

\section{Methods}

 The primary research question of the current project is: to what extent can open source data and tools be used to forecast eviction filings in Bronx County, NY? We ask this question in order to build the evidence base for the use of open source data and tools in human service intervention development. Below, we chronicle the process by which we gathered and analyzed the various open source tools and data.  
 
\subsection{Case study context}

To investigate the applicability of open source data and tools for human service intervention development, this paper uses  open source tools and data to build an analytic strategy to forecast eviction trends in Bronx County, New York. Specifically, we ask whether and how open source data and tools can support forecasting evictions during the economic shocks caused by the COVID-19 pandemic. We focus on non-payment eviction filings because we assume that any pandemic-related housing instability is driven by economic and financial strain. The Bronx is employed as a case study locale because it was shown to be an epicenter of COVID-19 infection rates in New York City \parencite{Thompson2020}, as well as an epicenter of the foreclosure crisis in New York City between 2008-2013 \parencite{Chan2011}. For example, in the months after COVID-19, residential non-payment eviction filing rates plummeted as a result of the temporary eviction moratoria put into effect at federal and state levels (\parencite[see]{evictionlab_NY} as well as figure 1). Further, the Bronx collectively experienced a 77.6\% decrease in residential non-payment eviction filings from March 2020 – June 2021 (i.e. during the moratoria) as compared to March 2018 – June 2019. Together, these  trends suggest a lower resourced area: the promise of open source tools and data include the ability to leverage little resources in order to scale technical solutions \parencite{chopra2008decoding}. In our analysis, we attempt to leverage open data and tools to provide a range of forecasts on the effect that COVID-19 had on housing instability in the Bronx. \emph{We conduct these analysis from the position of a potential provider invested in the community and aiming to assess immediate housing need.} 



\subsection{Data} 

Data for the current analyses came from three open sources. First, we gathered data on eviction filings from January 2016 through July 2021 from the Housing Data Coalition’s court filing records , which in turn compiled their data from the New York State Office of Court Administration  \footnote{The Housing Data Coalition can be found at \url{https://www.housingdatanyc.org/}; the NYS OCA can be found at \url{https://ww2.nycourts.gov/apps/chrs/index.shtml}}. Eviction filing data from OCA is available daily at the zip code level; we refer to this dataset as the OCA court filing data below. Second, we gathered data on employment and demographics for zip codes in the Bronx from 2016-2019 from the U.S. Census Bureau’s American Community Survey 5-year estimates \footnote{ACS 5 year estimates can be found at \url{https://www.census.gov/programs-surveys/acs.html}}. ACS estimates occur yearly. Finally, we used data from the DeepMaps Model of the Labor Force \parencite{ghitza2020deep} which provided monthly estimates of demographic data at the census tract level in 2020.\emph{Census tract data from DeepMaps were aggregated up to the zip code level using the Department of Housing and Urban Development's Census Tract to ZIP crosswalk file, which provides the ratio of residential addresses in each census tract that can be allocated to different zip codes} \parencite{din2021hud}. \footnote{It is important to note here that the Deep Maps model, data and website, was developed by TSD Communications \url{https://www.tsd.biz/} and Catalist ventures \url{https://catalist.us/}, two for profit data mining companies that made this particular data freely available. This is a point we will return to during the discussion.}  The reader should note that zip code 10464, representing Pelham Bay Park, was excluded from these analyses: 10464 was noted as an extreme outlier due to having a small number of renter-occupied housing units and subject to very few eviction filings.

\subsection{Independent Variables and Descriptive Statistics}

The first stage of data analysis is establishing descriptive statistics. Table 1 below offers the sociodemographic characteristics of Bronx County, and compares it to national averages. These data were derived from the ACS 2019 estimates, apart from COVID-19 case rates, which were compiled from New York City’s Department of Health and the Centers for Disease Control. 

\begin{quote}
\centering
    [Insert Table 1 here]
\end{quote}

The primary independent variables of interest used throughout these analyses and forecast models is the percentage of a given zip code reporting as “non-White” and the percent of the population 16 years and older who reported being employed (i.e., the employment-population ratio). For the years 2016-2019, the independent variables are derived from ACS 5-year estimates. In the current analyses, “non-White” is operationalized as individuals who identify in any category other than non-Hispanic White. This choice reflects a key limitation of open data which we will discuss in more depth in the discussion section:  granularity.  We use the employment-population ratio because it is a widely used economic indicator that considers both labor force participation and unemployment \parencite{donovan2015reclaiming}. This is especially useful when examining the economic impacts of COVID-19, as many individuals who are out of work are not actively searching for jobs due to stay-at-home orders, concerns for their health and safety, caring for children who are home from school, or because the scarcity of work opportunities has discouraged the continuation of job searching \parencite{bluedorn2021gender, coibion2020labor}. 

For 2020, we use the monthly estimates of employment and demographic characteristics available from the DeepMaps Model of the Labor Force, which compiles data from the Current Population Survey (CPS) microdata, the American Community Survey, and the Local Area Unemployment Statistical Program (LAUS). While these data sources provide the most up-to-date employment indicators available, they offer us another limitation of open data: temporal and geographic granularity. The DeepMaps Model of the Labor force employs modified multilevel regression and poststratification (MRP) methods to help overcome these limitations. The benefit of this approach is that it provides monthly demographic and employment estimates at the census-tract level in a timely manner, which is critical for researchers responding to current events such as COVID-19. In a final limitation of open data which we will discuss later in this paper, the Deep Maps model has not continued its offerings of data estimates. 
	
Given the open data and tools at our disposal, our analytic plan required that we use the DeepMaps data to estimate how changes in the employment-population ratio during the COVID-19 pandemic contributed to housing instability across communities in the Bronx. Since the OCA housing court filings data were available at the zip code level, we used zip codes as our unit of analysis. Accordingly, we aggregated tract-level DeepMaps demographic and employment variables “up” to zip codes based on the crosswalk files and recommended methodology outlined by the Department of Housing and Urban Development’s Office of Policy and Development and Research, \emph{which also accounts for cases where Census Tracts are split by zip codes} \parencite{din2021hud}.

\subsection{Dependent Variables}

The primary outcomes for the current analyses are (a) the total number of residential non-payment eviction filings in the Bronx and (b) the number of monthly residential non-payment eviction filings per 1,000 renter-occupied housing units (hereafter referred to as filing rate). We base our analyses on residential non-payment eviction filings because we assume the impacts of COVID-19 on housing instability operate via economic and financial strains that negatively impact household's ability to pay rent. Non-payment filings in the Bronx represent about 86\% of all residential eviction filings in the OCA filings data from January 2016 through July 2021.
	
Each instance of a non-payment eviction filing in the OCA filings data was accompanied by a filing date and the zip code where the property was located, which we then used to aggregate monthly filings per zip code. To allow for comparisons across zip codes in the Bronx, monthly eviction filings were standardized as a rate per 1,000 renter-occupied housing units in each zip code for models presented in the Analysis section below. The number of renter-occupied housing units in each zip code was acquired from the ACS 5-year estimates for each corresponding year (2016-2019). Figure 1 shows the number of monthly eviction filings in the Bronx between January 2016 and \textbf{July 2021 (the end of our forecast window)}. Figure 2 shows the average number of monthly eviction filings per 1,000 renter-occupied housing units for each zip code in the Bronx from 2016-2019. In Figure 2, the number of renter-occupied housing units was taken as the average number of renter-occupied housing units between 2016-2019.
\begin{quote}
\centering
[Insert figure 1 here] 
\end{quote}

\begin{quote}
\centering
[Insert figure 2 here]
\end{quote}

\subsection{Analysis} 

 Below we detail the process by which we analyzed the data. To be sure, this process is detailed such that it would require a human service agency to have at least one staff member capable of reasonably advanced statistics. We chronicle the analysis not in order of the figures and tables offered, but in the order that analysis unfolds due to three main caveats of open source data. First, while any data set one uses for analysis will require 'cleaning' - e.x. combining redundant variables or removing/imputing missing values - open data usually also require scaling to ensure the level of measurement is consistent throughout the dataset. For example, it may be the case that certain variables are at the block level, but we need them at the zip code level. Second, as one engages in the process of ensuring data are a the correct level of measurement, one may need to engage in more than one analytic conversion. Lastly, in the event a level of measurement is unavailable, the analyst may need to use a proxy variable to come closer to the outcome of interest. Recall that open data are generally made available due to policy requirements: as such, there are stipulations as to what information can be shared. Thus, what follows is an analysis that attempts to forecast evictions in Bronx County centering marginalized populations (in this case, we assume marginalized individuals and groups are communities of color), but does so using  meaningfully constrained open data. 

The analysis began using a traditional process of descriptive statistics. Specifically, we use Pearson’s correlations and linear regression models to establish the relationship between the percent of a zip code reporting as non-White and the decrease in the employment-population ratio during the onset of the pandemic from February 2020 - May 2020 (these are given in Table 2). We chose this time frame to capture pandemic-related decreases in the employment-population ratio as these months represent the minimum and maximum values for employment-population ratio in the DeepMaps data (see Appendix A). Then, we used a series of models trained on pre-COVID (January 2016 - December 2019) eviction filings and Census data to forecast the monthly eviction filing rate for each zip code in the Bronx in the months after the onset of the pandemic (i.e., our forecast period, January 2020 - July 2021). This allowed us to predict eviction rates and trends across the Bronx \emph{had the temporary eviction moratorium not been put into place during our forecast period} (see Figure 3).  

Having concluded the descriptive statistics, we move on to describing the construction of the forecast models. We employ multilevel models (MLMs) found in the R package \texttt{lme4} \parencite{bates2015fitting} in order to account for the clustering of our independent variables (employment-population ratio within each zip code, the percent of each zip code reporting as non-White, and time) within zip code. This process occurred in to two steps. First,  MLMs trained on pre-COVID data were run as a series of nested models to evaluate the extent to which time, the employment-population ratio, and percent of the population reporting non-White, improved model fit over an intercept-only model with a random effect at the zip code level. These nested models are displayed in Table 3. Nested models were then compared  to each other using Likelihood Ratio Tests as well as AIC and BIC values to gauge model fit. The random-intercept model (i.e., a random effect at the zip code level) which also included time, employment-population ratio, and percent of a zip code reporting as non-White was determined to be the best fit: \emph{this model is referred to as MLM 3 in Table 3.}
	
Having found the best model fit, we then used the parameters from our pre-COVID MLMs to forecast monthly eviction filing rates throughout the onset of COVID-19 and through our forecast period (January 2020 through July 2021) for each zip code in the Bronx. we do this using monthly demographic and employment estimates from the DeepMaps data: this analysis allows us to simulate the impact of COVID-19 on eviction files in the absence of policy intervention. However, here we hit another important feature of open data. Because the DeepMaps data are  only available through December 2020, projections of monthly eviction filings rates during 2021 assume the employment-population ratio recovered at a rate consistent with that seen occurring the DeepMaps data - between May 2020 and December 2020 (See Appendix A). Because of this important data caveat, we needed to find a way to update the employment population-ratio using only what we had available. Our solutions was to forecast each zip code’s employment-population ratio from January 2020 to July 2021 using the May 2020 to December 2020 DeepMaps data. We do this by specifying a random-intercept, random-slope MLM using time as the only independent variable.  In order to keep this statistical slight of hand conservative, we capped forecast employment-population ratios at their pre-COVID peak (February 2020; see Appendix A). Finally, we make separate forecasts using both raw DeepMaps employment-population ratios and a more conservative, adjusted value that aligns the DeepMaps January 2020 estimates with the 2019 ACS Census data (the unadjusted and adjusted values are included in Appendix A, respectively). The results from all MLM models can be found in Table 3.

In order to compare MLM3 (or, the final best fit model) with what actually happened during this time period, we engage in one more analysis. Specifically, we specify an Error, Trend, and Seasonality (ETS) model, using the "fpp3" package in R \parencite{fpp3}, as described by \parencite{hyndman2018forecasting}. In forecasting analysis, ETS models are used to capture time trends and seasonality effects. These models can does this while using exponential smoothing methods to assign weighted averages to observations in the training data. Importantly the technique stipulates the weights diminish as observations become older (i.e., more recent data has a greater effect on the model). In our analysis, the ETS model provided estimates of the eviction filing rate during our forecast period (January 2020 - July 2021) based only on eviction filings in the past (January 2016 - December 2019). As a final note, we used a “bottom-up” approach to reconcile the ETS model projections, where each zip code was forecast individually and then aggregated “up” to give estimates of Bronx County as a whole \parencite{hyndman2018forecasting}.
	
\section{Results}

Before describing the analysis results, we begin by highlighting the various assumptions we found to be a prerequisite of working with the open source data in this study. First, the analysis assumes the relationship between the employment-population ratio, race, and monthly eviction filing rates did not fundamentally change due to COVID-19 or other exogeneous factors not captured by our models. Second, the analysis assumes that the number of renter-occupied housing units remained stable from 2019 through the end of our forecast window.  Third, the analysis assumes that non-payment eviction filings are a proxy for economic-related housing instability. Fourth, the analysis assumes that the employment and demographic variables between the ACS and DeepMaps data are comparable in how they are being captured. As discussed in the Methods section as well as Appendix A, we acknowledge these possible differences by making forecasts with both unadjusted and adjusted DeepMaps employment-population ratios. Given the assumptions needed, it is perhaps not surprising that our models lead to over-estimation, which we discuss below.

\subsection{Establishing Which Sections of the Community Might Require Support}


\begin{quote}
\centering
[Insert Figure 3 here]
\end{quote}

\begin{quote}
\centering
[Insert Table 2 here]
\end{quote}


Recall that our primary analytic aim is to take the perspective of a human service provider and forecast failure-to-pay evictions in the community we serve. Assuming the Bronx is the community we serve, then the first thing we want to do is establish the non-payment eviction trends immediately following the event that precipitated our need to forecast. Immediately following the onset of Covid-19, we find that residential non-payment eviction filings dropped along with the employment-population ratio. Across the zip codes in the Bronx, there was an 11.8 percentage point decrease (from February 2020 to May 2020) in the population 16 years and older who were employed. Across the Bronx, these decreases ranged from 7.4 points in zip code 10471 to 14.2 points in zip code 10458 (see Figure 2). Linear regression and Pearson correlation tests did not reveal a significant association between the percent of a zip code reporting as non-White and the employment-population ratio in February 2020 (i.e., the month prior to the onset of COVID). However, race was significantly associated with the percentage point decrease in the employment-population ratio from February 2020 to May 2020 (B = 0.09, CI [0.06, 0.12], p < 0.001, see Table 2). This association remained after controlling for the employment-population ratio in February 2020 (see Table 2). In other words, according to the estimates provided by the DeepMaps models, communities with higher proportions of people of color likely saw the largest decreases in the number of employed  people during the onset of the pandemic. From the perspective of a human service provider, that let's us know that this subpopulation might be most in need of supportive services across a continuum of care (i.e. from financial resources to socio-emotional support). 

\subsection{Establishing Non-Payment Eviction Filing Trends before Covid-19}


\begin{quote}
\centering
[Insert Table 3 here]
\end{quote}

Having established the parts of our community which have been hit hardest by the pandemic, we turn now to establishing non-payment eviction filing trends before the pandemic. Doing this allows us to see what the percentage change in eviction filing trends is using the same data sources. That is, though we may have information from other public sources concerning eviction filing trends, using open source data and tools requires us to be sure we are comparing apples to apples within our analysis. Table 3 presents the results of a series of nested multilevel regression models predicting the average number of monthly non-payment eviction filings per 1,000 renter-occupied housing units from 2016-2019 (i.e., our pre-COVID training period). A random-intercept model including time, the percent of zip code who were employed, and the percentage of a zip code who reported being non-White proved to have the best fit (Table 3, MLM 3). This model suggests the average monthly eviction rate between 2016-2019 dropped steadily across this time period (B = -0.13, \text{CI} [-0.15, -0.10], p < 0.001).  The employment-population ratio of a zip code was significantly negatively associated with the average monthly eviction filing rate (B = -0.27, \text{CI} [-0.52, -0.01], p < 0.05). That is, as the employment-population ratio of a zip code increases, the eviction filing rate decreases: perhaps unsurprisingly, the relationship between employment and non-payment eviction is strong. Lastly, the percentage of the population within a zip code reporting as non-White had a positive significant association with the average monthly eviction filing rate (B = 0.22, \text{CI} [0.14, 0.29], p < 0.001). That is, as the percentage of non-white persons increases in these zip codes, so too does the eviction filing rate, for this time period. From a human service perspective, these data would suggest that human services in the Bronx should be tailored to the communities of color who reside there and be oriented towards  housing and employment issues. Importantly, these results are about the 3 years \emph{just before the pandemic}.  

\subsection{Forecasting Non-payment Eviction Filing trends from January 2020 – July 2021}

Having established the relationship between our outcomes of interest and the variables we think help explain them, we now turn to forecasting the expected monthly eviction filing rate, using the publicly available DeepMaps Model of the Labor force as our input data. The forecasting frame is January 2020, the first month that granular employment data is available via DeepMaps, through July 2021, the month that the federal eviction moratorium was set to expire. We make three forecasts based on the perviously described MLMs. The first forecast used the time-only model (MLM 1). This model proved a superior fit to the intercept-only model (MLM 0) and captures the time trend of the eviction rates steadily decreasing from 2016-2019. While this trend may not persist into the future, incorporating this time trend ultimately gives a more conservative estimate of the number of evictions we would expect to see post-COVID.  Other MLM forecasts incorporated the employment-population ratio and the percentage of a zip code’s population reporting as non-White. In addition to the three forecasts based on the previous MLMs, we also forecast the monthly non-payment eviction filing rate using a bottom up ETS model (Table 3). Figure 3 shows the forecast monthly eviction rate per 1,000 renter-occupied housing units across zip codes in the Bronx based on our adjusted MLM 3. The figure illustrates how the sudden drop in the employment-population ratio at the onset of the pandemic, which was most pronounced in communities of color, had the effect of increasing the forecast eviction filing rate across zip codes in the Bronx. From a human services perspective, then, these data suggest the pandemic \emph{should have found us doing much of the same work as before: supporting communities of color with employment and housing resources.  }

\begin{quote}
\centering
[Insert Figure 3 here]
\end{quote}

Finally, we use the forecasted eviction rates from each model to calculate the number of cumulative evictions we would expect to see during our forecast window between January 2020 through June 2021. These calculations were made by taking the monthly forecast eviction rate in each zip code and multiplying it by the number of renter-occupied housing units in 2019. Table 4 below illustrates how our time-trend only forecasts (i.e., MLM 1 and ETS Bottom-Up) give very similar estimates for the number of cumulative eviction filings expected (85,185 vs. 81,966). The models which incorporate employment-population ratio and the percent of the zip code reporting as non-White (i.e., MLM 3 and MLM 3 adjusted) forecast a greater number of evictions than our time-trend only models. For example, our MLM 3 model predicted 20\% more eviction filings than did our time trend only MLM 1 and our adjusted MLM 3 model predicted 11\% more eviction filings than our MLM 1 time trend only model during during our forecast period. 
                    [Insert Table 4 here]
                    
The results of this analysis, from the perspective of human services, offers a few key takeaways concerning the use of open source data and tools for intervention development. First, while incorporating demographic and economic indicators yields higher estimates of eviction filings than time-trend only models, all models presented here predicted evictions filings well above the actual number of filings that occurred during our forecast period of January 2020 through June 2021 (see Table 4). \emph{These models predicted between 2.62 and 3.26 times the number of eviction filings than were actually filed during this period}. Translated into practical terms, humans service professionals and organizations with the capacity to undertake these open source analyses would have been well prepared to advocate  for the community by being able to tell the potential story: many people would have suffered economic hardships beyond what was felt absent the eviction moratoria.  It is, however, not impossible for some of the estimation to be as a result of the use of open data. However, we argue the positive side of any potential overestimation in need may be  the potential for human services to be over prepared. Likely, if there is any  overestimation it is likley due to a key issue with open source data: professionals can only use what is included in publicly shared files.

\section{Discussion and Conclusion}

The current work sought to investigate the extent to which open source data and tools can be used to forecast eviction filings in Bronx County, NY. Our team undertook this endeavor as a result of concerns during the Covid-19 pandemic regarding the potential impact of evictions on homelessness. We employ open source data from ACS, the DeepMaps Model of the Labor Force, as well as the Housing Data Coalition’s compilation of the New York State Office of Court Administration (OCA) eviction filing records. We use the open source statistical software tool R to run descriptive analyses on eviction filing trends in Bronx County, NY before Covid-19. We use the same tool to then train various multilevel models to forecast eviction filing trends immediately following the onset of Covid-19. The forecasts presented here estimate the number of eviction filings during Covid-19 to have been between 2.62 and 3.26 times the actual number that occurred. Given the analytic assumptions made, and the analytic checks on the impact of those assumption found in the appendices, we conclude that eviction filing trends in Bronx County NY would have been significantly greater if not for the policy interventions that occurred during that time frame. \textbf{These findings suggest public data can be useful in estimating the need of local communities.} This, in turn, encourages future research investigating effective and efficient methods to train human service professionals in data science techniques, which includes proficiency in open data and open source analytic tools \parencite{rodriguez_introducing_2024, Romich2021}. 

This investigation contributes to the literature on open source data and tools in two concrete ways. First, the manuscript demonstrates the use of both in a human service need forecasting analysis. Because the data and tools are open source, the entire analysis can be replicated with the data and code found here \url{https://anonymous.4open.science/r/open_source_and_tools_2025/README.md}. Second,  this manuscript offers ana analysis conducted using a transdiciplinary approach, a cornerstone of the open source movement. The authors come from a variety of fields including social work, mathematics, computer science, and computational social science. As a team, they leverage over 50 collective years of practice experience within the fields of housing policy, affordable housing development, as well as mental and public health. The wealth of information each member brought to the team resulted in a project in service to the public interest, which we hope will be replicated across the human services research community. Further, our collaboration made use of a medley of approaches to  support the use of open, though imperfect, data sources to effectively assess a local level resource need. In addition to these benefits, our analysis also offers key lessons in the use of open data and analytics tools for human service intervention development. 

\subsection{Open Data and Human Service Intervention Development}

While our findings suggest open source data are at least a reasonable approach to analyses attempting to forecast human service needs for intervention development, they are not without drawbacks. First, our analysis shows that relying on open data for human service intervention development necessitates acquiring data from multiple sources. For example, the current analysis has 3 source of data, one of which (the housing coalition data) is itself comprised of two sources. Interested human service researchers and professionals should make note that \emph{the veracity of open data are impacted by the number of hands they have to pass through to become publicly available}. However, we argue the use of open data and tools outweighs this potential drawback because \emph{open data and tools allow for research and program development that is by default, at least more transparent than research with pay-walled data} \parencite{Jaakola_etal_2014}. For example, in the case of housing specifically, data are extremely difficult to come by. From non-existent data sources to incomplete counts of required data to astronomical data acquisition fees, housing research on national housing trends (and thereby intervention development) remains exceedingly difficult absent a centralized system of data collection. Further, calls for centralized data collection necessarily require increased surveillance on American households and we hesitate to join such calls without clear ethical guidelines for collection, storage, and fair use. In the meantime, the above manuscript offers a proof of concept for real-time service need forecasting with data and tools available to anyone. 

Second, the analyses presented in this paper illustrates how evictions may be better predicted by accounting for a greater variety of demographic characteristics, which we argue facilitates a targeted approach to intervention development. Though the findings are presented from a case study of one county, our results suggest projections of potential resource needs, and thereby the allocation of resources, should consider how economic and health shocks exacerbate the drivers of human service needs above and beyond what can be captured by historical trends alone. Projecting the magnitude of resource need without considering health and economic shocks is likely to lead to underestimation: our research suggests \emph{these underestimations may be disproportionately large among marginalized communities}. Therefore, we argue future research attempting to forecast human resource needs would do well to remain bound to geographic areas operating under the same resource policies. Binding forecasts to local areas also supports intervention development by allowing for contextualized policy recommendations. 

However, the granularity of open source data is an issue for human service organizations. Specifically, while census block level data is perhaps the most useful for human service need projection, open data at this level is difficult to access. For open data that can be converted to block level, data compilers (be they individuals or organizations) may have a difficult time keeping such datasets open source across time. For example, the Deep Maps data used in the current analysis was created and released by 2 private data broker firms, 1 of which is tied to lobbying efforts. These facts raise a variety of questions: 

\begin{enumerate}
    \item how  did these companies collect these data? 
    \item how does the provenance of the data impact estimates?
    \item why are these data not available from other (public) sources?
    \item does usage of these open data translate to endorsement of other activities that such data may have facilitated? 
\end{enumerate}

While one might laud a for-profit company making data public, they are under no obligation to disclose their data gathering process. Of course, if such well-intentioned data brokers followed the open source definition and protocols described by \parencite{niu_survey_2021, opensource_initiative, debian} these questions would be answered more satisfactorily. As of the writing of this manuscript, however, the data from Deep Maps cannot be used without returning to the two proprietary entities to ask them for the source code. Further, though the data for our time period is openly accessible through their website, the collected data have not been updated. This lack of maintenance is one of the often-seen problems of open datasets \parencite{gebru2021datasheets}. 

\subsection{Open Analytic Tools and Human Service Intervention Development}

 Within the National Association of Social Workers (NASW) code of ethics - the professional body under which many human service professional operate - ethical standard 1.13(a), states 
 
\begin{quote}
    " When setting fees, social workers should ensure that the fees are fair, reasonable, and commensurate with the services performed. Consideration should be given to clients’ ability to pay" \parencite{NASW2021Ethics}.
\end{quote}

We argue this ethical standard would be well applied within human services research, \emph{especially when that research is community facing}.  We define community facing as work that is about or with any individual or group, especially in circumstances when that work is tied to a human service organization. Unlike other forms of social scientific inquiry, human service research is accountable to people who use and provide human services. \emph{Recognizing that human service organizations likely cannot pay the thousand dollar licensing fees associated with mainstream statistical software packages like SAS and STATA, the use of open source software is as much a matter of justice as it is a matter of staying technological relevant.} 

Using open source analytical tools would support human service research in two other ways. First, \emph{open source analytic tools facilitates the open dissemination of research}. Bowen and colleagues \parencite{bowen_open_2013} argue increasing open access publishing is not only in line with the aforementioned NASW code of ethics, but also supports reproducible science within the field. The use of open source statistical software removes paywall barriers to the reproduction of research, a crucial requirement in any community based intervention development. Reproducible research has arguable cascading benefits for human service research: increasing human service scholar's ability to develop and sustain interdisciplinary partnerships, as well as facilitating the consumption of human services research by more fields. Increasing the reproducibility of human services research may allow support increasing the problems that human services scholars can help solve.

\clearpage

\printbibliography

\end{document}